\documentclass[aps,pra,twocolumn,groupedaddress]{revtex4}

\usepackage{graphicx}

\begin{document}

%\preprint{}

\title{Transient dynamics and momentum redistribution in cold atoms \\ via recoil-induced resonances }


\author{Joel A. Greenberg}
\email[]{JAG27@phy.duke.edu}
\author{Daniel J. Gauthier}
%\thanks{}
%\altaffiliation{Duke University Physics Department}
\affiliation{Department of Physics,  Center for Nonlinear and Complex Systems, and  Fitzpatrick Institute for Photonics, Duke University, Durham, North Carolina 27708, USA}


\date{\today}

\begin{abstract}
We use an optically dense, anisotropic magneto-optical trap to study recoil-induced resonances (RIRs) in the transient, high-gain regime.  We find that two distinct mechanisms govern the atomic dynamics:  the finite, frequency-dependent atomic response time, and momentum-space population redistribution.   At low input probe intensities, the residual Doppler width of the atoms, combined with the finite atomic response time, result in a linear, transient hysteretic effect that modifies the locations, widths, and magnitudes of the resulting gain spectra depending on the sign of the scan chirp.  When larger intensities (\textit{i.e.}, greater than a few $\mu$W/cm$^2$) are incident on the atomic sample for several $\mu$s, hole-burning in the atomic sample's momentum distribution leads to a coherent population redistribution that persists for approximately 100 $\mu$s.  We propose using RIRs to engineer the atomic momentum distribution to enhance the nonlinear atom-photon coupling. We present a numerical model, and compare the calculated and experimental results to verify our interpretation.

\end{abstract}

%\keywords{}

\maketitle

\section{Introduction}
\label{sec:intro}

Much recent work has focused on the realization of nonlinear optical interactions with few photons for application to creating more efficient all-optical components and to quantum information and communication schemes \cite{lukin03, chang07, dawes2008}. In order to drive a material into the nonlinear regime with only a small number of photons, the nonlinear material must interact strongly with the incident radiation. An optically  thick atomic sample can provide strong atom-photon coupling, but the sample must be prepared in such a way that the deleterious effects of linear absorption are mitigated.  To date, most of the techniques that have been proposed to increase the atom-photon interaction strength (\textit{e.g.}, EIT, cavity QED) only rely on manipulating internal- or cavity-atomic states.  By combining previously-used techniques with new approaches that use control of both internal and center-of-mass atomic states, novel methods for realizing enhanced nonlinear optical interactions can be achieved \cite{gupta2007}.  Furthermore, because these methods primarily involve center-of-mass atomic states, they are widely-applicable to a broad range of atomic species and less sensitive to optical and magnetic field inhomogeneities than quantum-interference-based schemes.

In this paper, we report on an  approach that exploits the collective excitation of a spatially extended, optically thick sample of cold atoms. Specifically, we focus on using a phenomenon known as recoil-induced resonance (RIR), which can be viewed in terms of Raman transitions between the quantized momentum states of an atom \cite{guo92,courtois94}.   For a given pump-probe detuning, this two-photon process coherently transfers atoms from one resonant momentum state to another, thus coupling the internal and external center-of-mass atomic states.  The atom-photon coupling strength directly depends on the populations of atoms in the initial and final momentum states.  Thus, by selectively engineering the instantaneous atomic momentum distribution \cite{olshanii2000}, we can enhance the coupling strength and reduce the threshold for nonlinear optical behavior.

In order to understand how to construct an optimal momentum distribution via RIRs, we must first study the transient dynamics of RIRs in the high-gain regime.   For weak optical fields incident on optically thin atomic samples,  momentum-changing RIR events alter negligibly the momentum distribution of the atoms.  Thus, for a thermal gas of atoms with a Maxwell-Boltzmann momentum distribution, the resulting RIR signal has a Gaussian-derivative shape with small gain (loss) for negative (positive) pump-probe detunings.  For an optically thick atomic sample in the high-gain regime, the RIR feature dominates the observed spectrum \cite{veng2005a}.  In order to understand the  complex, highly-coupled dynamics in this case, one must consider the interplay between the amplification of the probe beam and modification of the momentum distribution as the field propagates through the gas of atoms.  Furthermore, one must consider the finite response time of the material when investigating transient phenomena.

As an example of self-enforced momentum engineering, Vengalattore \textit{et al.} claim to observe enhanced nonlinear optical effects at low light levels giving rise to transient  optical bistability  \cite{veng08}.  Here, we report a similar, transient hysteretic effect at low light levels ($<10$ $\mu$W/cm$^2$), but interpret it as a purely linear phenomena.  Also, we observe a substantial modification of the atomic momentum distribution at higher intensities, thus demonstrating the feasibility of momentum-state engineering via RIRs.  We compare the experimentally-obtained results with a numerical model to verify our interpretation of the observations.

   This paper is organized as follows.  In Section \ref{sec:exp}, we briefly describe the experimental setup and  Section \ref{sec:theory} describes the model used to describe the RIRs. We present and discuss the results in Section \ref{sec:results} and Section \ref{sec:discussion}, respectively.  Section \ref{sec:conclusion} concludes the paper and indicates future research directions.

\section{Experimental Setup}
\label{sec:exp}
In the present study, we create a sample of cold atoms via an anisotropic magneto-optic trap (MOT) as described in previous work \cite{greenberg07}.  The sample consists of approximately $5\times10^{8}$ $^{87}$Rb atoms confined in a cylindrical volume with a $1/e$ radius and length of 300 $\mu$m and 3 cm, respectively.  A pair of counterpropagating lin$\bot$lin laser beams (cooling beams) intersect the trapping volume (as defined by the magnetic field) at a small angle ($\theta\sim10^{\circ}$) to cool the atoms along the long dimension of the trap;  we achieve typical atomic temperatures of $20-30$ $\mu$K with this scheme. These cooling beams also act as the pump beams in the RIR scheme, and have an effective Rabi frequency of $\Omega_1/\Gamma=2.5$ (where $\Gamma/2\pi=6$ MHz is the natural linewidth of the $5S_{1/2}(F=1) \leftrightarrow 5P_{3/2}(F=2$) transition).  The experiment is run in steady-state (with the MOT beams on), and all of the beams are typically detuned $\Delta=3-5$ $\Gamma$ below the atomic resonance.  This configuration enables the production of optical densities (OD; $I_{out}/I_{in}=\textnormal{exp}(-OD)$) of up to $\sim 60$.  We control the OD by varying the detuning of the repump laser beam, which allows us to investigate RIRs in both the low- and high-gain regimes.

\begin{figure}
  \includegraphics[width=3.375 in]{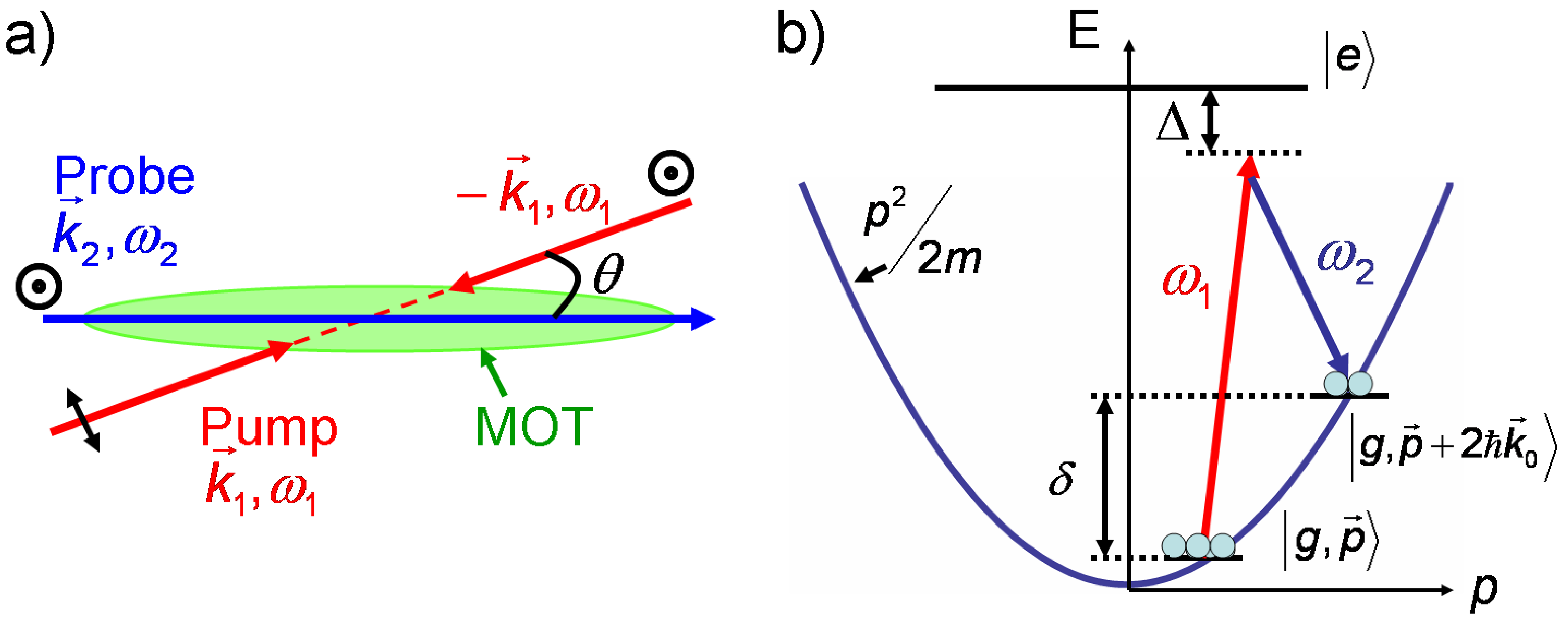}
  \caption{a) Experimental RIR beam geometry.  A pair of counterpropagating, lin$\bot$lin beams, oriented at an angle $\theta$ with respect to the long trap direction, act as cooling beams for the MOT and as pump beams in the RIR scheme.  The probe beam propagates along the length of the trap, and is copolarized with the counterpropagating pump beam.  b) RIR energy level scheme. The horizontal and vertical axes correspond to the atomic momentum and energy, respectively, and the parabola describes the quantized ladder of states that satisfy energy and momentum conservation. The circles represent the  relative populations of the two resonant momentum states. }
  \label{fig:setup}
\end{figure}

 The probe beam used for RIR spectroscopy is split off from the pump beam, which provides the phase coherence necessary for studying spectroscopically narrow, multiphoton resonances.  As shown in Figure \ref{fig:setup} a), the probe beam propagates along the long axis of the trap, with a polarization that is parallel to that of the counterpropagating pump beam (\textit{i.e.}, the $\aleph\bot$ configuration described in Ref. \cite{guo94}).The frequency detuning of the probe beam relative to the pump beam and spectral scan rate of the probe beam are independently controlled via acousto-optic modulators. The input probe intensity ($I_{in}$) ranges from 0.5-200 $\mu$W/cm$^2$, and the scan rates vary between 0.08-10 MHz/ms.  The size of the incident probe beam is $100$ $\mu$m ($1/e^2$ intensity radius) with a Rayleigh range longer than the trap length.  Although independent experiments by our group \cite{greenberg08a} and others \cite{veng05b} have demonstrated  waveguiding and focusing/defocusing effects that play additional roles in light propagating through such a dispersive atomic medium, we do not explicitly consider such effects here.

\section{Theoretical Description}
\label{sec:theory}
Most formulations of RIRs consider the atom-photon interaction as a stationary process, where the predicted spectrum is independent of the prior history of the field or atomic ensemble \cite{berman99}.   This approach is appropriate for weak incident fields and slow probe frequency scan rates (relative to the atomic decay times), but one must solve the transient problem in the case of fast scan rates or substantial probe beam amplification.  Theoretical \cite{guibal96}  and experimental \cite{kozuma95} studies of RIRs in the transient regime have been carried out in the low-gain limit for optically thin atomic samples. While the majority of studies focusing on high-gain center-of-mass-mediated phenomena have focused on the collective atomic recoil lasing (CARL) regime \cite{bonifacio94,bonifacio97}, some recent work has also focused on the RIR regime \cite{veng08}.

In order to describe the present experimental situation, we use a model that describes the interaction of classical optical fields with a sample of thermal, cold atoms with quantized momentum states \cite{moore98}. Experimentally, we find that the main contribution to the RIR spectrum comes from the probe beam and nearly counterpropagating pump beam (for the polarization configuration shown in Fig. \ref{fig:setup}), and hence we only consider these two beams in the theory.  Furthermore, because the angle between the pump and probe beams is small and the atoms are tightly confined in the radial direction by the trapping potential, we consider only motion along the longitudinal direction of the trap.  Finally, we consider that the atoms have two internal states (a ground state $\mid g \rangle$ and excited state $\mid e \rangle$) coupled to a quantized ladder of momentum states (see Fig. \ref{fig:setup} b).

We write the relevant Hamiltonian  as \cite{moore98}
%\begin{widetext}
\begin{eqnarray}
\mathcal{H}&=&\sum_k \Big[\frac{\hbar^2k^2}{2m}\hat{c}_g^*(k)\hat{c}_g(k)+\left(\frac{\hbar^2 k^2}{2 m}+\hbar \omega_0\right)\hat{c}_e^*(k)\hat{c}_e(k) \nonumber \\
 &&+i \hbar \sum_{j=1,2} (g_j a^*_j e^{i \omega_j t} \hat{c}_g^*(k-k_j)\hat{c}_e(k)-H.c.)\Big] ,
\end{eqnarray}
%\end{widetext}
where $m$ is the atomic mass and $\omega_0$ is the natural frequency of the two-level atomic transition.  The atom-photon coupling constant for the pump ($\vec{k_1}, \omega_1$) and probe ($\vec{k}_2\sim-\vec{k}_1, \omega_2=\omega_1-\delta$) beams are given by $g_{1,2}$, where $g_j=\mu_j[c k_j/(2\hbar \epsilon_0 V)]^{1/2}$, $\mu_j$ is the dipole matrix element, and $V$ is the quantization (trap) volume.  The unitless, single-photon field amplitudes of the pump (probe) fields are $a_1$ ($a_2$), and $\hat{c}^*_{e,g}(k)$ ($\hat{c}_{e,g}(k)$) are creation (annihilation) operators for the ground and excited states with atomic momentum $\hbar k$, respectively.

In the limit that the pump  and probe
 beams are far-detuned from the
atomic resonance, we can  adiabatically eliminate  the excited
states. Assuming that the pump beam propagates with negligible attenuation,
one can derive an expression for the coherence between any pair
of initial  and final momenta.  By considering only
the populations ($\Pi_p~=\rho(p,p)$) of  and
the first-order coherences ($\eta_p=\rho(p+1,p)e^{-i\delta t}$) between momentum states, the atomic evolution
can be described as \cite{veng08}
\begin{eqnarray}
\dot{\Pi}_p&=&[-i\beta^*a_2(-\eta_p+\eta_{p-1})+c.c.]-\gamma_{pop}(\Pi_p-\Pi_{th,p}), \nonumber \\
\dot{\eta}_p&=&i[4\omega_r(p^2-(p+1)^2)-\delta(t)+i \gamma_{coh}]\eta_p \nonumber \\
&& -i\beta a_2^*(\Pi_{p+1}-\Pi_p),
\label{eqn:main1}
\end{eqnarray}
where $\gamma_{pop}$ ($\gamma_{coh}$) are the population
(coherence) decay rates, $\Pi_{th,p}$ is the thermal population
distribution (typically given by a Maxwell-Boltzmann distribution),
and $\omega_r=\hbar k_1^2/2m $ is the single-photon recoil frequency. The dimensionless momentum is given by
$p=\hbar k/(2\hbar k_0)$ (for $2k_0=|\vec{k_1}-\vec{k_2}|$), and $\beta=g_1 g_2 a_1/\Delta$, where $\Delta=\omega_2-\omega_0$ is the pump-bare atomic resonance detuning (see Fig. \ref{fig:setup} b).

In order to present a self-consistent picture of the atom-field interaction, Maxwell's equations must also be solved simultaneously.
Ignoring propagation effects (\textit{i.e.}, using a mean-field approximation), the time dependent probe field can be
written as:
\begin{equation}
\dot{a_2}=-\frac{\kappa}{2}(a_2-a_{in})+i\beta N \sum_p\eta^*_{p-1},
\label{eqn:main2}
\end{equation}
where $N$ is the number of atoms in the probe beam volume, $\kappa=c/L$ is the free space
decay of photons from the atomic sample of length $L$,  $a_{in}$
is the amplitude of the input probe beam, and the summation runs over all momentum states.

We numerically integrate Eqs. \ref{eqn:main1}-\ref{eqn:main2} using a spacing between momentum states of $10^{-2}p$ (we have verified that the results do not change when we use smaller steps), and then sum over momenta in the range $p=[-35,35]$. The gain experienced
by the probe beam propagating through a distance $\Delta z$ is
calculated as
\begin{equation}
I_{out}/I_{in}=\textnormal{exp}[2((a_2-a_{in})/a_{in})(\Delta z/L)].
\end{equation}
 This formulation allows us to model the rich, coupled dynamics that we describe in Sec. \ref{sec:results}.   Before investigating the full solution to Eqs. \ref{eqn:main1}-\ref{eqn:main2}, we briefly discuss two limiting cases: the thermal-equilibrium limit and the perturbative limit.

 \subsection{Thermal-Equilibrium Limit (TEL)}
 We first consider the situation of arbitrary input probe beam powers, but we fix the atomic momentum distribution at its thermal equilibrium value ($\Pi_p(t)=\Pi_{th,p}$).  In this limit, excitations between the atoms (through the coherence $\eta_p$) and the field (through $a_2$) can be exchanged. This limit allows us to distinguish between effects produced by population redistribution and those due to either the finite response time of the material or the fact that the atomic sample consists of an inhomogeneously-broadened group of radiators.  Because all of these mechanisms can give rise to similar transient effects, we compare both the numerical results from the full set of evolution equations and the fixed-population, thermal-equilibrium limit equations with the experimental results in  Sec. \ref{sec:results}.

 \subsection{Perturbative Limit}
 In order to connect our results with prior work \cite{guibal96,guo92}, we investigate Eqs. \ref{eqn:main1}-\ref{eqn:main2} in the perturbative limit.  Here, we consider only weak beams and low gain, where the probe amplitude and amplification of the probe beam are assumed to be small (\textit{i.e.} $a_2(t)\sim a_{in}<<a_1)$.  Also, as in the TEL, we assume that the momentum distribution remains at its thermal equilibrium  value ($\Pi_p(t)=\Pi_{th,p}$).  Under these approximations, Eqs. \ref{eqn:main1}-\ref{eqn:main2} reduce to
 \begin{equation}
    \dot{\eta_p}=i[f(p)-\delta(t)]\eta_p-\gamma_{coh}\eta_p-i \beta a_{in}^*(\Delta\Pi_p),
     %\dot{eta_p}=(-4i\omega_r(2p+1+\delta(t))-\gamma_{coh})\eta_p-i \beta \a_in^*(\Pi_{th,p+1}-\Pi_{th,p}).
 \end{equation}
where $f(p)\equiv 4\omega_r(p^2-(p+1)^2)$ and $\Delta\Pi_p \equiv \Pi_{th,p+1}-\Pi_{th,p}$.  If $\delta(t)=\delta_0+R t$, where $\delta_0$ is the probe detuning at $t=0$ and $R$ is the probe scan rate, then $\eta_p(t)$ can be explicitly solved for as
\begin{eqnarray}
\eta_p(t)&=&\left(\frac{2 i \pi}{R}\right)^{1/2}(\Delta\Pi_p)\textnormal{exp}\left[-\frac{i}{2R}(\delta_{eff}+R t)^2\right]\times \nonumber\\
&&[\textnormal{erfi}(\sqrt{\frac{2i}{R}}\delta_{eff})-\textnormal{erfi}(\sqrt{\frac{2i}{R}}(\delta_{eff}-R t))],
\label{eqn:pertlim}
\end{eqnarray}
where $\textnormal{erfi} \equiv -i \textnormal{erf}(iz)$ is the imaginary error function and $\delta_{eff}=\delta_0-f(p)-i\gamma_{coh}$ is the effective, complex probe  detuning from the momentum class $p$.  %We can then calculate the amplitude of the resulting probe beam via Eqn. \ref{eqn:pertlim}, as described above in Eqn. \ref{eqn:main2}.
 Equation \ref{eqn:pertlim} can describe both the transient behavior for a given detuning (for $R=0$) and the RIR spectra obtained by scanning the probe beam ($R\neq0$). While we do not rely on this model for quantitative comparison between theory and experiment, having an analytical solution allows us to gain insight into the physical mechanisms involved.

\section{Results}
\label{sec:results}

\subsection{Transient Dynamics}
\label{sec:transdyn}

In order to understand the interplay between the finite response time of the atoms and the redistribution of atomic population among momentum states, we study RIRs over a range of input probe intensities and probe frequency scan rates. We begin by investigating the temporal evolution of the probe beam amplitude for a fixed pump-probe detuning.  We note that, for the case of $R=0$ in the perturbative regime (\textit{i.e.}, constant probe frequency), Eq. \ref{eqn:pertlim} reduces to the result for the RIR transition rate per atom with momentum $p$ ($dP_p/dt$)  obtained by Guibal \textit{et al.} \cite{guibal96}
\begin{equation}
\frac{dP_p}{dt}\propto \Im[\eta_p]=(\Delta\Pi_p)\frac{\sin(\delta_{eff}t)}{\delta_{eff}},
\label{eqn:pertfixed}
\end{equation}
where $P_p$ is the probability of a Raman transition between a state with momentum $p$ and $p+2k_0$.  After integrating over all momentum classes, the perturbative limit predicts that the probe amplitude initially increases (decreases) for a negative (positive) detuning, and that it oscillates at a frequency of  $\delta/\pi$ before decaying to its steady-state value. One can interpret this result physically by considering the individual momentum classes as independent oscillators.  As Eq. \ref{eqn:pertfixed} shows, each momentum class has a characteristic coherence oscillation time that depends on the pump-probe detuning.  Thus, the probe field oscillations arise due to the interference of the radiation emitted by the various momentum classes, and the decay occurs as a result of net destructive interference as the oscillators dephase relative to one another.

   \begin{figure}[htb]
  \includegraphics[width=3.375 in]{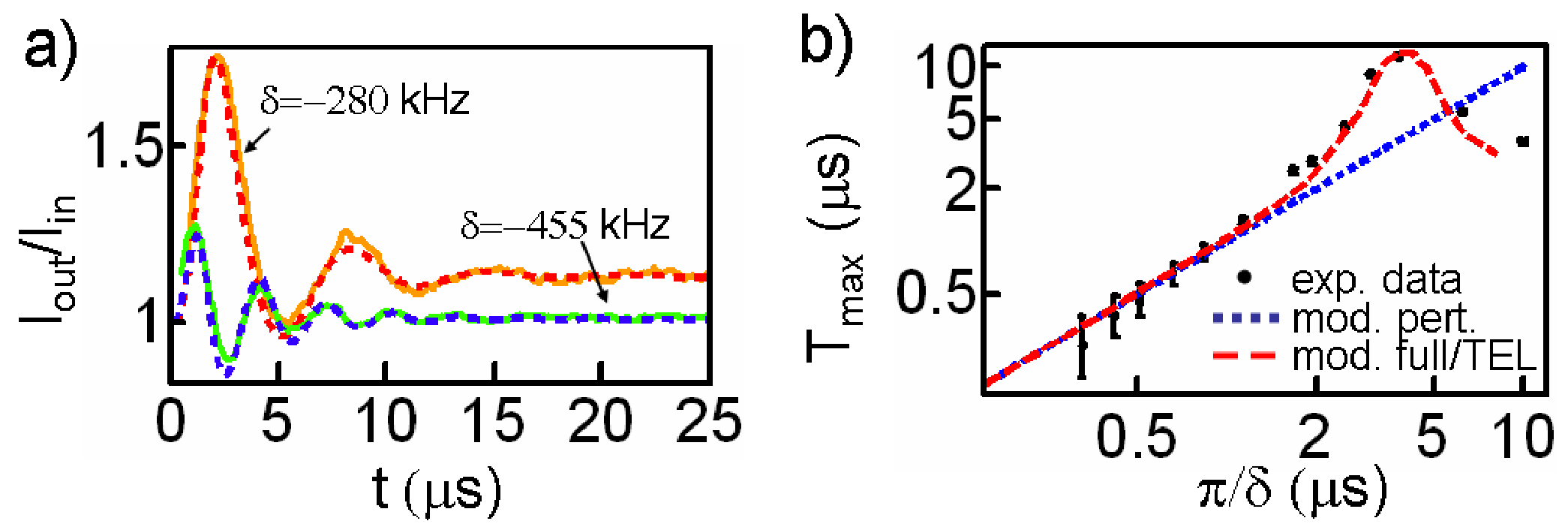}
  \caption{a) Transient  RIR oscillations observed when the probe beam is turned on. Experimental data (solid lines) is compared with the numerical results (dashed lines) of the full model for $\delta=-280$ and $-455$ kHz.  We normalize the magnitude of the experimental signal to facilitate comparison with theory. b) Time at which the first probe maxima occurs as a function of the inverse detuning. Other experimental parameters are: $I_{in}=100$ $\mu$W/cm$^2$, $T=20$ $\mu$K, $N=8\times10^6$, $\beta=13$ kHz. }
  \label{fig:offon}
\end{figure}

   To investigate the transient regime experimentally, we first produce a cloud of atoms, and then rapidly turn on the probe beam and measure its intensity after it passes through the atoms ($I_{out}$).  Figure \ref{fig:offon} a) shows the experimentally-observed probe intensity along with numerical calculations from the full equations for two different detunings.  We note that, while the oscillation frequency depends only on the pump-probe detuning, the decay time is sensitive to the probe power, average atomic temperature, and the coherence decay rate. By independently measuring the atomic temperature (via in-situ RIR velocimetry \cite{meacher94}) and probe power, we determine $\gamma_{coh}/2\pi=20$ kHz for a variety of pump-probe detunings by fitting the data to the full model.

   In general, we see good agreement between the  numerical and experimental results over a range of detunings and probe powers.  Figure \ref{fig:offon} b) shows that, at short times, the first maxima occur at a time $T_{max}$, which is approximately equal to $\pi/\delta$, as predicted by all three models.  For longer times, the experimentally-observed values of $T_{max}$ diverge from the perturbative solution.  The results of the TEL and full-equation solutions, though, agree throughout all investigated times, thus indicating that, over such short times (several $\mu$s), population redistribution does not affect the dynamics.

\begin{figure}[htb]
  \includegraphics[width=3.375 in]{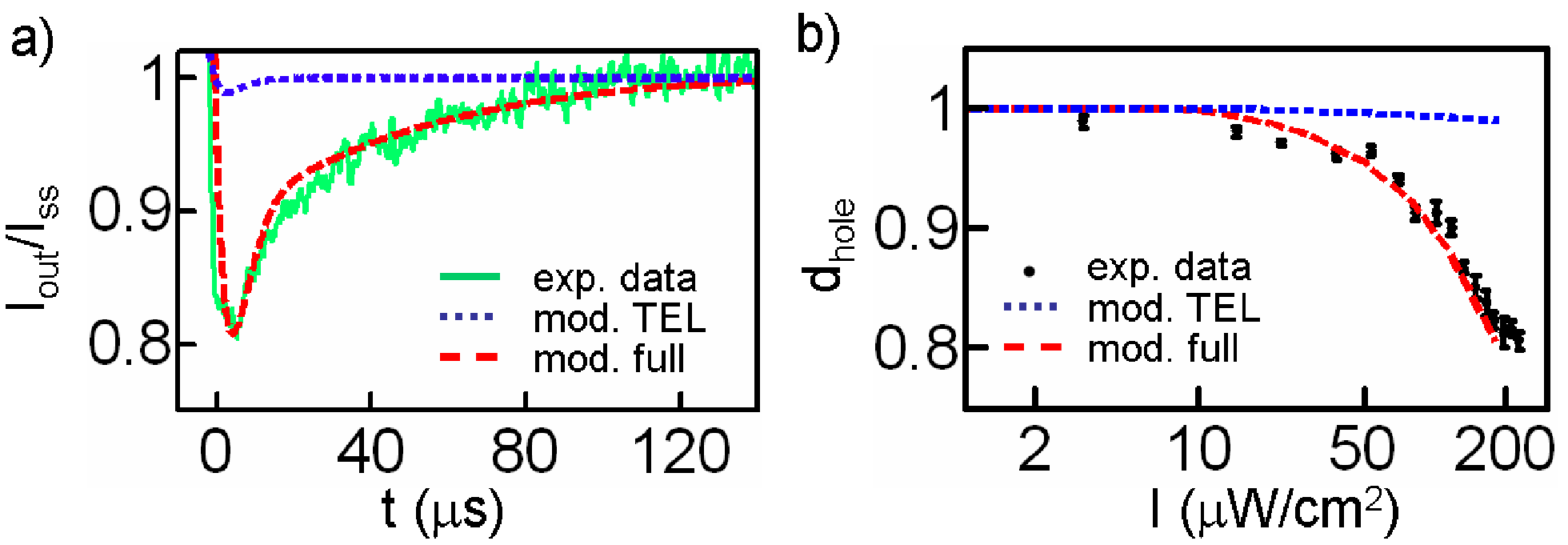}
  \caption{a) Experimentally-observed and numerically-predicted temporal evolution of the probe intensity occuring after the probe intensity is switched from 200 to 1 $\mu$W/cm$^2$. b) Relative hole depth resulting from different initial probe beam intensities. Other experimental conditions are: $T=20$ $\mu$K, $N=8\times10^6$, $\beta=13$ kHz, and $\delta=-140$ kHz. }
  \label{fig:highlow}
\end{figure}

At slightly longer times, the effects of population redistribution and rethermalization become important. In order to study this regime, we performed an experiment similar to Ref. \cite{veng08}.  We illuminate the atoms with an intense probe beam for $\sim100$ ms before rapidly dropping the probe beam intensity to 1 $\mu$W/cm$^2$.  For large initial intensities, we measure small, weak-beam gains that slowly increase to their steady-state values.  To properly interpret this response, we need to distinguish between two different effects: the initial, rapid oscillations in the output probe intensity (as described in the previous paragraph) and the slower response due to changes in the momentum-space populations. Figure \ref{fig:highlow} shows that, even when the momentum distribution is fixed, we expect a small decrease in gain after we reduce the probe intensity.  This occurs because of the destructive interference between adjacent momentum classes excited by the strong beam that relax at different rates. In agreement with Vengalattore \textit{et al.}, we interpret the slower relaxation to steady state as due to a redistribution of atomic population.  The intense beam removes atomic population from a particular momentum class, thereby reducing the gain; once the probe intensity is reduced, random scattering and collisions rethermalize the atoms, thus refilling the hole burned in the momentum distribution and increasing the gain.

We estimate  the time necessary to remove a substantial amount of population for given an incident power as $t_{pump} \sim 2\pi / \Omega_{12}$ (where $\Omega_{12}=4 \beta a_2$ is the two-photon Rabi frequency).  For typical experimental parameters  (\textit{i.e.}, $\beta=13$ kHz and a beam diameter of 100 $\mu$m), the time required to redistribute population is given as
\begin{equation}
t_{pump}[\mu s]\sim167 \times (I_{in}[\mu \textnormal{W/cm}^2])^{-1/2}.
\label{eqn:tpump}
\end{equation}
 Equation \ref{eqn:tpump} predicts a population redistribution time of $t_{pump}=53$ $\mu$s for an intensity of $I_{in}=10$ $\mu$W/cm$^2$, corresponding to an input energy density of $\sim$2 photons/($\lambda^2/2\pi)$.  In the following paragraphs, we compare these predictions with the numerical and experimental results.

Figure \ref{fig:highlow} a) shows a representative plot of the normalized probe gain ($I_{out}/I_{ss}$, where $I_{ss}$ is the steady-state probe intensity) as a function of time  for an initial probe intensity of 200~$\mu$W/cm$^2$.  By fitting the measured signal with the full experimental model, we extract a rethermalization rate of $\gamma_{pop}/2\pi= 3.4$ kHz.  We repeat this experiment for a range of initial probe intensities, and observe that the  relative hole depth ($d_{hole}=I_{min}/I_{ss}$, where $I_{min}$ is the minimum transient gain) decreases with the initial intensity.  Furthermore, $d_{hole}$ is equal to unity for intensities less than  $\sim10$ $\mu$W/cm$^2$, indicating that momentum-space population redistribution negligibly effects the atom-field dynamics for sufficiently low intensities.

To verify our interpretation of the temporal evolution of the gain, we compare the numerical results from the full and TEL calculations with the experimental data.  Figure \ref{fig:highlow} b) shows the values of $d_{hole}$ as a function of the strong probe beam intensity. The TEL predicts a slight reduction in gain due to transient effects identical to those described above and does not agree with the data. On the other hand, the full solutions accurately predict both the relative hole depth and decay time.   Thus, in agreement with Eq. \ref{eqn:tpump}, we find that population redistribution plays an important role when intensities greater than a few $\mu$W/cm$^2$ are incident upon the atoms for  tens of $\mu s$.

\subsection{Transient Hysteresis}

We also investigate the effects of momentum redistribution by studying the RIR spectrum.  We measure the spectrum by scanning the probe beam frequency across the resonance at a fixed rate $R$.  While, in steady state, the resulting RIR signal reflects the derivative of the equilibrium momentum distribution, scan rates that traverse the resonance on time scales comparable to the RIR turn-on dynamics ($\sim1$ $\mu$s) produce history-dependent spectra that reflect the local momentum distribution. In this section, we focus on the effects of the frequency-dependent, finite response time of the material and the controllable redistribution of momentum-space population on the RIR spectra.

\begin{figure}[htb]
\centering \includegraphics[width=3.375 in]{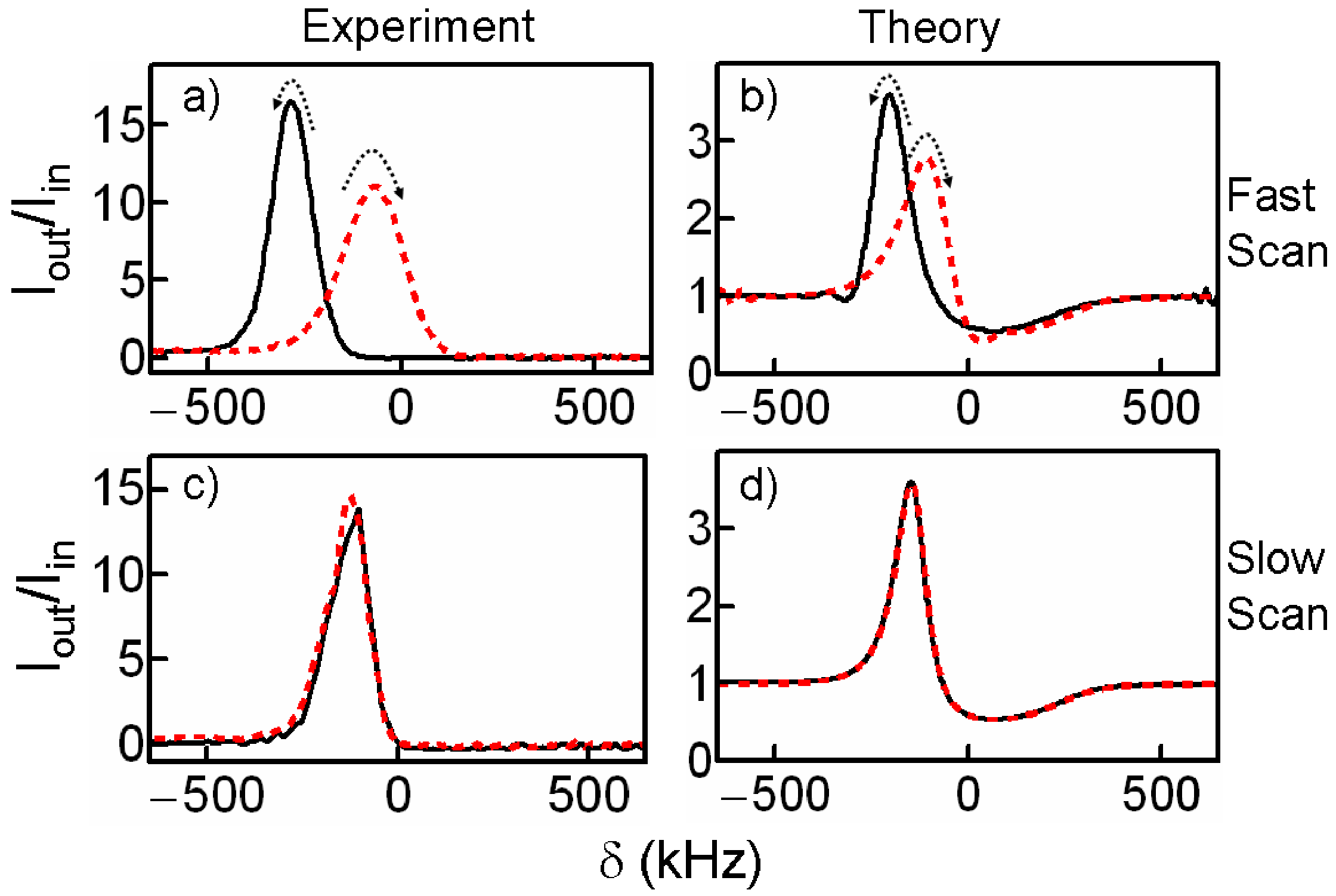}
\caption{ a), c) Experimental and b), d) theoretical ($T=20$ $\mu$K, $N=8\times10^6$, $\beta=13$ kHz) RIR
spectra at slow and fast scan rates, respectively, for a probe intensity of 0.5 $\mu$W/cm$^2$. Solid (dashed) lines correspond to negative (positive) chirps.}
\label{fig:lowpower}
\end{figure}

Figure \ref{fig:lowpower} shows the RIR spectra for a low input probe intensity (0.5 $\mu$W/cm$^2$) at different scan rates.  For slow scans,  positive ($R>0$) and negative ($R<0$) chirps produce identical spectra, whereas fast scans result in resonance line shapes with altered locations of the maximum value of the gain, widths, and amplitudes. Figure \ref{fig:lowpower} a) and c) show experimental spectra obtained with a slow ($R=0.1$ MHz/ms) and fast ($R=8$ MHz/ms) scan rate, respectively.   The slow scan probes the steady-state response of the sample, and we observe no chirp dependence.  For a fast scan with a negative chirp, though, the spectrum has a larger gain peak, narrower gain feature, and a larger shift from $\delta=0$ relative to the steady-state case. For a positive chirp, the opposite is true. The chirp-dependence of the numerically-calculated spectra, shown in Fig.  \ref{fig:lowpower} b) and d), agree well with the experimental data.  We note, though, that the amplitude of the gains predicted by the model do not match  the experimentally-measured values.  We believe that this discrepancy is caused by two main effects not accounted for in the model:  additional resonances caused by the MOT beams (producing small, additional Raman resonances between light-shifted levels \cite{brzozowski05,chen01}), and propagation effects that the mean-field assumption ignores \cite{veng05b, bonifacio97}.

One can understand this transient hysteresis in terms of the short-time dynamics of the atomic sample. We first note that, because the momentum distribution is not significantly modified at very low probe powers applied for short times (see Sec. \ref{sec:transdyn}), one cannot explain the results of Fig. \ref{fig:lowpower} in terms of population redistribution.   Rather, the hysteretic effect occurs as a result of the frequency-dependent response time of the atoms.  As mentioned above, the time it takes for a particular momentum class to reach its first maximum is approximately inversely proportional to the pump-probe detuning.  When the probe frequency is scanned farther from $\delta=0$, the resonant momentum classes respond increasingly quickly as $\delta$ increases, resulting in a situation where a range of momentum classes radiate at their maximum intensity at the same time.  In the opposite case, when the probe frequency is scanned toward $\delta=0$, the resonant momentum classes reach their maximum radiated powers at different times.  Considering the resulting probe intensity as a superposition of this inhomogeneous collection of radiators thus explains the increased (decreased) and narrowed (broadened) gain peak for the fast, negatively (positively) chirped case.  In a similar way, the finite response time of the material effectively delays the occurrence of the gain peak for either case of the chirp, resulting in the observed shifts in the gain peaks.

\begin{figure}[htb]
  \includegraphics[width=3.375 in]{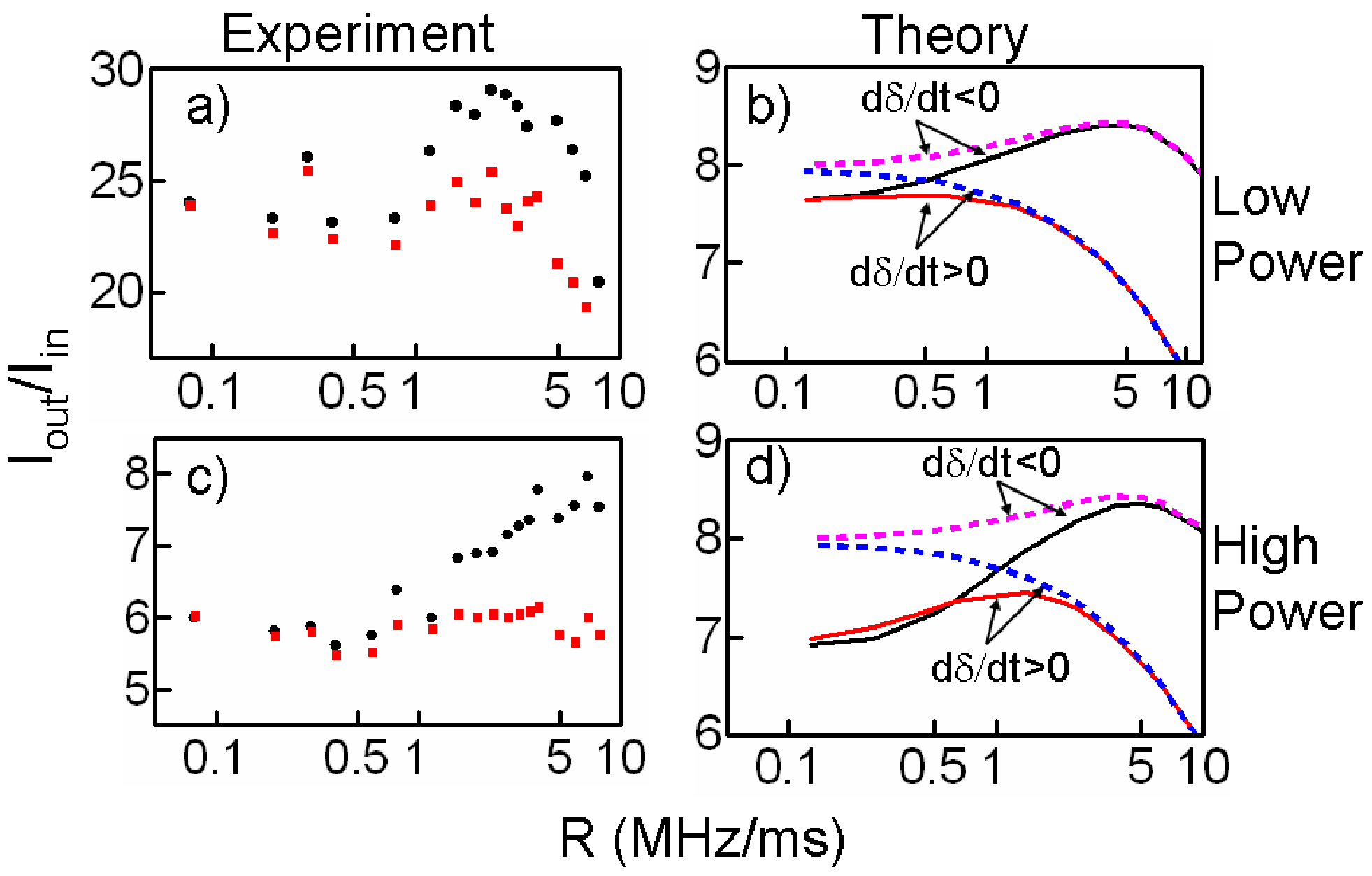}
    \caption{a) and c) Experimental results for the RIR peak gain as a function of scan rate for an input intensity of 0.5 and 100 $\mu$W/cm$^2$, respectively.  Boxes and circles correspond to positive and negative chirps.  b) and d) Theoretical results for the RIR peak gain as a function of scan rate for an input intensity of 0.5 and 100 $\mu$W/cm$^2$, respectively. Solid (dashed) lines correspond to solutions to the full (TEL) equations for positive and negative chirps, as indicated.  Other experimental conditions are: $T=20$ $\mu$K, $N=8\times10^6$ and $\beta=13$ kHz.}
  \label{fig:gvsrate}
\end{figure}

We confirm our interpretation of the hysteresis at low powers by first studying the perturbative solution given in Eq. \ref{eqn:pertlim}.  For finite scan rates, Eq. \ref{eqn:pertlim}  produces the experimentally-observed, chirp-dependent variations of the RIR gain feature.  As this model explicitly ignores changes to the momentum distribution as well as any back-action between the atoms and photons, one can understand the observed hysteresis as a transient, linear effect. This result is directly analogous to the modification of the resonance lineshape of a damped, driven harmonic oscillator when the driving frequency is scanned across the resonance.

To quantify the importance of population redistribution in the experimentally-observed spectra, we compare the results of the full and TEL calculations.  Figure \ref{fig:gvsrate} b) shows that both calculations predict almost the same chirp-dependence of the gain, thus indicating that   momentum redistribution does not significantly contribute to the observed gain.  Figure \ref{fig:gvsrate} a) shows experimental data that demonstrates that, for slow scans ($R<0.5$ MHz/ms), the atoms reach steady-state and the chirp-dependence disappears.  At faster scan rates, the gain increases, as discussed above. Beyond  $R\sim3$ MHz/ms, though, the gain observed for both chirps decreases as the atom-photon interaction time becomes too short.

At higher powers, population redistribution plays a role in the observed RIR spectra.  Figure \ref{fig:gvsrate} c) and d) show the measured and calculated dependence of the peak gain value on the scan rate for an input probe intensity of 100 $\mu$W/cm$^2$.    For fast scan rates, the numerical results of the thermal and full model coincide, thus indicating that the probe beam does not spend long enough at each detuning to significantly modify the populations.  For slow scan rates, the results of the two models diverge. In this region, momentum redistribution decreases the predicted gain by effectively reducing the population difference between the resonant momentum states.  This situation differs from Ref. \cite{veng08} because the scan rate considered here is slow enough to allow for new, equilibrium momentum distributions to occur for each resonant momentum class

\begin{figure}[htb]
  \includegraphics[width=3.375 in]{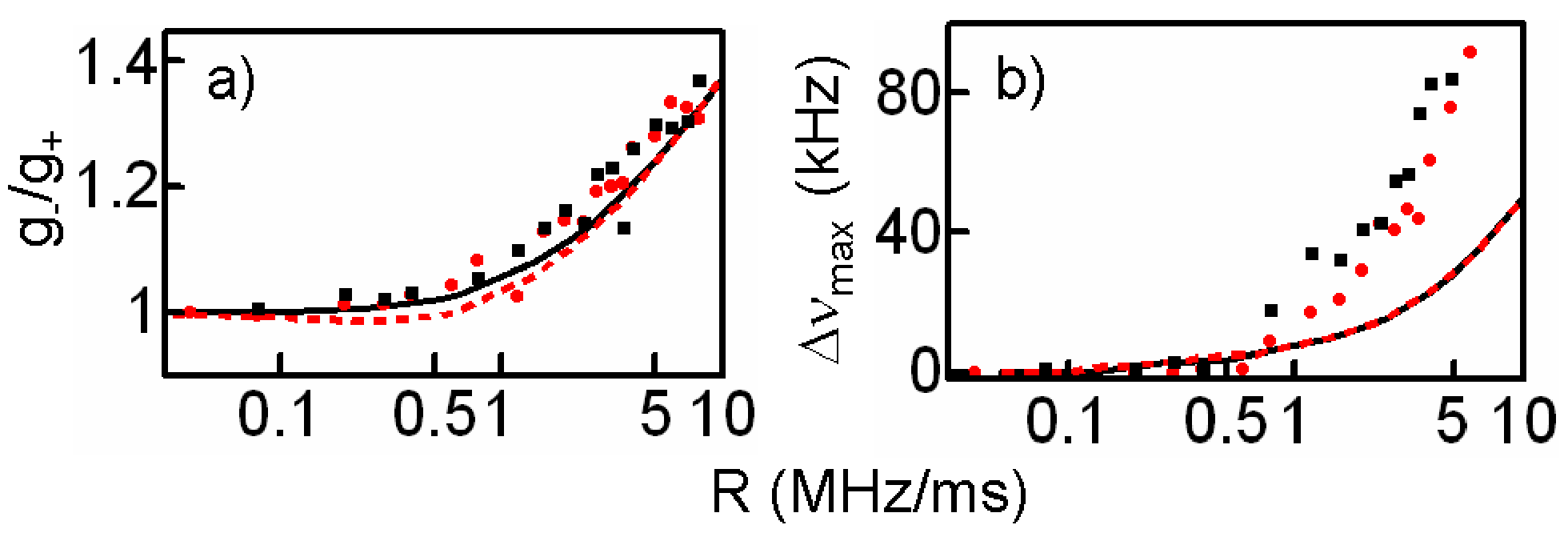}
  \caption{a) RIR peak gain and b) shift of the peak gain location as a function of scan rate.  Boxes and circles correspond to experimental results, and the solid and dashed lines correspond to theoretical results for input intensities of 0.5  and 100 $\mu$W/cm$^2$, respectively. Other experimental conditions are: $T=20$ $\mu$K, $N=8\times10^6$, $\beta=13$ kHz, and $\delta=-140$ kHz.}
  \label{fig:ratehigh}
\end{figure}

  The relative differences between the peak gain magnitude and location help further clarify the situation.  Figure \ref{fig:ratehigh} a) and b) show that the ratio of the negative ($g_-$) and positive ($g_+$) chirp peak gains and the relative shift of the location of the gain peak ($\Delta\nu_{max}=\delta_+^{max}-\delta_-^{max}$) agree qualitatively with results of the full model and do not strongly depend on intensity.  Thus, while population redistribution between momentum classes occurs in this system, the time scales over which it acts are longer than the time the probe spends resonant with a given momentum class for fast scans.  We therefore conclude that the observed transient hysteresis results from the linear, frequency-dependent response time of the inhomogeneously-broadened system, rather than coherent population redistribution.

 \section{Discussion}
 \label{sec:discussion}
  Our results are consistent with those described by Vengalattore \textit{et al.} \cite{veng08}; only our interpretation of the observed phenomenon differs. For the mechanism described by Vengalattore \textit{et al.}, a negatively-chirped scan shuffles population such that the population in the ground momentum state increases, thus increasing the population difference between the resonant momentum states and enhancing the gain.  For a positive chirp, though, one should observe a similar enhancement as the population difference is similarly increased by a removal of population from the excited momentum state.  As we observe only an increase in gain for a negative chirp, in agreement with the predictions of the perturbative model, we conclude that the effect described in  Ref. \cite{veng08} does not play a major role in our system.  Nevertheless, we clearly demonstrate effects of momentum redistribution.

\section{Conclusion}
\label{sec:conclusion}
In conclusion, we have investigated RIRs in the transient, high-gain regime, which supplements the work described in Ref. \cite{guibal96}.  By studying the RIR signal produced at short times for fixed frequencies, we measure the effective population and coherence decay rates.  Also, we note two important effects that influence the RIR spectrum: the frequency-dependent, finite response time of the material, and the redistribution of momentum-space population for sufficiently high probe intensities.  By measuring the RIR spectrum for various powers and probe frequency scan rates, we observe transient hysteretic phenomena that arise from both linear  (finite material response time)  and nonlinear (population redistribution) effects.  The results of this study demonstrate that momentum-space hole-burning is possible because of the inhomogeneously-broadened nature of the RIR, and that it persists for approximately 100 $\mu$s.  By tailoring the atomic momentum distribution via coherent population redistribution mediated by RIRs, the nonlinear atom-photon coupling can be controlled to enhance or reduce the nonlinearity.  Also, optimizing the setup to make use of collective effects resulting from atomic bunching in position space can further improve the observed nonlinear coupling.  Together, these effects make this system an excellent candidate for the realization of ultra-low-light nonlinear optics.

\begin{acknowledgments}
 We gratefully acknowledge the financial support of
the DARPA DSO Slow-Light Program.
\end{acknowledgments}

%\bibliography{RIR_paper}

\end{document}